\documentclass[12pt, letterpaper]{JHEP3}

\usepackage{amsmath}

\title{Enhanced Symmetries in Multiparameter Flux Vacua}

\author{Oliver DeWolfe\\
Department of Physics\\
Princeton University\\
Princeton, NJ 08544, U.S.A.\\
{\tt odewolfe@princeton.edu}}

\abstract{We give a construction of type IIB flux vacua with discrete R-symmetries and vanishing superpotential for hypersurfaces in weighted projective space with any number of moduli.  We find that the existence of such vacua for a given space depends on properties of the modular group, and for Fermat models can be determined solely by the weights of the projective space.  The periods of the geometry do not in general have arithmetic properties, but live in a vector space whose properties are vital to the construction.}

\preprint{hep-th/0506245, PUPT-2166}

\newcommand{\cp}{\mathbb{CP}}
\newcommand{\wcp}{\mathbb{WCP}}
\newcommand{\lcm}{{\rm lcm}}
\newcommand{\bZ}{\mathbb{Z}}
\newcommand{\bQ}{\mathbb{Q}}

\begin{document}
\section{Introduction}

Flux compactifications of string theory provide an interesting and increasingly well-studied avenue towards making contact with phenomenology.  Compactifications with flux can stabilize moduli, potentially eliminating one of the chief difficulties of traditional string theory backgrounds.
Although interesting progress in moduli stabilization via fluxes has been made in heterotic \cite{Heterotic}, M-theory \cite{MTheory} and most recently type IIA compactifications \cite{IIA}, the most well-studied case remains that of compactifications of type IIB string theory \cite{IIB}.  In type IIB string theory, fluxes can completely stabilize the complex structure moduli and the dilaton, while the K\"ahler moduli are unfixed; it has been argued that nonperturbative effects can then in some cases stabilze these remaining moduli \cite{KKLT, Nonpert}.

As there are apparently a large number of possible flux compactifications, one is naturally led to ask about the properties that distinguish them.  A particularly interesting characteristic to understand is the presence of {\it enhanced symmetries}.   Other than the natural interest in classifying the different types of low-energy theories making up the ``landscape" of vacua \cite{CosConst, Statistics}, enhanced symmetries are interesting for a number of reasons.  Discrete symmetries are useful for model-building in both phenomenological and cosmological contexts.  
In addition, it has been argued that the fraction of vacua with vanishing superpotential is associated to the prevalence of vacua with realistically low supersymmetry breaking scale \cite{DGT}; see also \cite{Scales}.  These vacua also have vanishing cosmological constant at tree level and hence fit onto the ``third branch" of \cite{DGT}, which was argued to have different statistics than other branches; this was further studied in \cite{DOS}.  Hence understanding how common such vacua are is of considerable interest.

In \cite{DGKT}, supersymmetric\footnote{Here we mean vacua solving the F-flatness conditions associated to the complex structure moduli and dilaton; when the superpotential vanishes these are supersymmetric at tree level.} flux vacua with vanishing (tree level) superpotential and/or discrete symmetries were studied in compactifications with no more than one complex structure modulus.  There it was found that enhanced symmetry vacua occured in some models but not others, though the associated geometries were naively similar.  Given that one knows these kinds of vacua are indeed present in some cases, it is clear that it would be desirable to understand how common they are in more general cases.

Moreover, the presence of such vacua seemed in \cite{DGKT} to be associated with a particular mathematical structure appearing in the periods characterizing the geometry: the periods took values in a finite-dimensional algebraic extension of the rational numbers, and this fact was important for obtaining a solution to the vacuum equations.  Especially considering the relevance of such arithmetic features of geometries to issues such as attractor vacua \cite{AA}, one is naturally led to wonder whether this sort of mathematical structure persists for more complicated models.  Other work on the arithmetic properties of Calabi-Yau geometries can be found in \cite{OtherGeom}.

In this paper, we address the question of the incidence and properties of enhanced symmetry vacua associated to geometries with more than one complex structure modulus.  We focus on a particular class of spaces, the hypersurfaces in weighted projective space, though we see no reason why our results should not generalize.  We look for constructions of the type of vacuum that was most interesting in \cite{DGKT}: vacua with vanishing superpotential enforced by a preserved discrete R-symmetry.
A complementary study, also searching for R-symmetric vacua in weighted projective spaces but using different methods, appears in \cite{DS}.

Although naively it seems the vacuum equations associated to the additional moduli will greatly complicate the analysis, we obtain a construction under which these equations become trivial.  We show that if certain constraints on the fluxes can be imposed, all vacuum equations follow and the $W=0$ vacua exist.  A consequence of this is that not all complex structure moduli are fixed at tree level.
Whether these constraints on the fluxes can hold is determined by the properties of the modular group transformations for a particular space, and for the simplest (Fermat) hypersurfaces, this can be determined {\it entirely} from the integer weights characterizing the projective space.  We find that, as for the one-parameter case, for a fixed number of moduli some but not all spaces can support $W=0$ vacua.  Estimating how many do brings in simple aspects of number theory, and one finds that in the most favorable cases, the number of $W=0$ vacua approaches the same scaling as the set of all vacua as the number of moduli becomes large.

Furthermore, we find that although the mathematical structure of the periods continues to play a vital role, in multiparameter models they generally {\it do not} live in an algebraic field extension of the rational numbers.  Instead, they generate a vector space with basis elements generated by in general transcendental numbers, without any natural product.  The properties of this vector space are critical to the existence of vacua.

In section~\ref{VacuaSec}, we review flux compactifications of type IIB string theory and the results of \cite{DGKT} on enhanced symmetry vacua.  In section~\ref{GeomSec} we review the relevant properties of the periods of hypersurfaces in weighted projective space.  Section~\ref{ConstructSec} describes our construction of enhanced symmetry vacua, while section~\ref{ExampleSec} applies the method to a number of examples, including one outside our class of spaces.  In section~\ref{ConclusionSec} we conclude.

\section{Review of enhanced symmetry vacua}
\label{VacuaSec}

\subsection{Type IIB flux compactifications}

We use the same conventions as \cite{DGKT}.  We work with a Calabi-Yau 3-fold ${\cal M}$ and choose a standard symplectic basis for the $b_3 = 2 (h_{2,1} + 1)$ 3-cycles $\{ A^a,  B_b \}$ with dual cohomology elements $\alpha_a, \beta^a$ obeying
\begin{eqnarray}
\int_{A^a} \alpha_b = \delta^a_b \,, \quad \int_{B_a} \beta^b = - \delta_a^b \,, \quad \int \alpha_a \wedge \beta^b = \delta_a^b \,.
\end{eqnarray}
The relevant moduli are the complex combination of the RR axion and the dilaton $\phi \equiv C_0 + i e^{-\varphi}$, as well as the complex structure moduli, which are encoded in the periods of the holomorphic 3-form $\Omega$,
\begin{eqnarray}
z^a = \int_{A^a} \Omega \,, \quad \quad {\cal G}_b = \int_{B_a} \Omega \,,
\end{eqnarray}
where the $z^a$ may be used as projective coordinates on the $h_{2,1}$-dimensional complex structure moduli space, with the ${\cal G}_b = \partial_b {\cal G}(z) $ taken as functions of the $z^a$.  Defining the $b_3$-vector of periods $\Pi(z) \equiv ({\cal G}_b, z^a)$, we may write the K\"ahler potential for the dilaton and the complex structure moduli as 
\begin{eqnarray}
\label{KahlerPot}
K &=& K^\phi + K^{cs} \,, \quad \quad K^\phi = - \log (-i (\phi - \bar\phi)) \,,  \\
\quad K^{cs} &=& - \log (i \int_{\cal M} \Omega \wedge \bar\Omega) = - \log(-i \Pi^\dagger \cdot \Sigma \cdot \Pi) \,,
\end{eqnarray}
where $\Sigma$ is the symplectic matrix $\Sigma \equiv \begin{pmatrix}\;0\;\;\; 1 \cr -1\;\; 0\end{pmatrix}$.

We turn on the RR and NSNS 3-form field strengths,
\begin{eqnarray}
\label{Fluxes}
F_{3} = - (2 \pi)^2 \alpha'(f_a\, \alpha^a + f_{a +h_{2,1}+1}\, \beta_a)
\,, \quad H_{3} = - (2 \pi)^2 \alpha'(h_a \, \alpha^a + h_{a
+h_{2,1}+1} \,\beta_a) \,, 
\end{eqnarray}
where flux quantization requires the $b_3$-vectors $f$, $h$ to have integer entries.  The fluxes generate a tadpole for the charge associated to the $C_4$ field, with value
\begin{eqnarray}
\label{Tadpole}
N_{\rm flux} = {1 \over (2 \pi)^4 ({\alpha'})^2} \int_{\cal M} F_3 \wedge H_3 = f \cdot \Sigma \cdot h \,,
\end{eqnarray}
which must be canceled against negative-charge sources such as O3-planes and $(p,q)$ 7-branes; in counting flux vacua one typically leaves this charge sink as an arbitrary integer $L$ and requires
\begin{eqnarray}
\label{LBound}
N_{\rm flux} \leq L \,,
\end{eqnarray}
where the case of an inequality can be made up by mobile D3-branes.  Most importantly, the fluxes (\ref{Fluxes}) induce a superpotential for the moduli $\phi, z^a$ given by \cite{GVW}
\begin{equation}
\label{Superpotential}
W = \int_{\cal M} G_3 \wedge \Omega = (2 \pi)^2 \alpha' \, (f - \phi h) \cdot \Pi \,, \quad \quad G_3 \equiv F_3 - \phi H_3 \,.
\end{equation}
In what follows we shall set $(2 \pi)^2 \alpha' = 1$.  The K\"ahler moduli $\rho_i$ do not participate in $W$, and as a result one may derive the no-scale relation
\begin{eqnarray}
\label{NoScale}
\sum_i |D_{\rho_i} W|^2 = 3 |W|^2 \,,
\end{eqnarray}
leading to the positive-definite potential
\begin{eqnarray}
V = e^K \left(|D_\phi W|^2 + \sum_a |D_{z^a} W|^2 \right) \,.
\end{eqnarray}
We shall neglect the K\"ahler moduli hereafter.

\subsection{$W=0$ vacua}

We are interested in vacua that satisfy the F-flatness conditions associated to the superpotential (\ref{Superpotential}),
\begin{eqnarray}
\label{DWZero}
D_{z^a} W \equiv \partial_{z^a} W + W \partial_{z^a} K = 0 \,, \quad \quad
D_\phi W \equiv \partial_\phi W + W \partial_\phi K = 0 \,, 
\end{eqnarray}
as well as the satisfying the vanishing of the superpotential,
\begin{eqnarray}
\label{WZero}
W = 0\,.
\end{eqnarray}
For solutions of (\ref{DWZero}), the tadpole $N_{\rm flux}$ (\ref{Tadpole}) becomes positive definite, and a finite number of vacua exist satisfying (\ref{LBound}).  As the charge sink $L$ becomes large, the number of such vacua scales as 
\begin{eqnarray}
\label{AllVacScaling}
N_{\rm vac} \sim \sqrt{L}^{2 b_3} \,,
\end{eqnarray}
where $2 b_3$ is the total number of fluxes.  One question is then how many vacua exist additionally satisfying (\ref{WZero}).  Vacua with $W=0$ are supersymmetric at tree level thanks to (\ref{NoScale}).

It is straightforward to show that (\ref{WZero}) combined with the dilaton equation from (\ref{DWZero}) is equivalent to
\begin{eqnarray}
\label{TwoEqns}
\int F_3 \wedge \Omega = \int H_3 \wedge \Omega = 0  \quad \Rightarrow \quad f \cdot \Pi(z^a) = h \cdot \Pi(z^a) = 0 \,,
\end{eqnarray}
where all dependence on the dilaton has dropped out.  Hence $W=0$ vacua may only exist at points on the complex structure moduli space where the period vector $\Pi$ is orthogonal to integral flux vectors $f$, $h$; moreover to avoid a vanishing tadpole (\ref{Tadpole}), the vectors $f$ and $h$ must not be aligned.  These periods must furthermore satisfy the remaining equations (\ref{DWZero}) for the complex structure moduli.

In \cite{DGKT}, examples of hypersurfaces in weighted projective space with a single complex structure modulus $\psi$ were considered.  The equations (\ref{TwoEqns}) were studied at the Landau-Ginzburg point $\psi = 0$, where the period vector $\Pi$ acquired a particular arithmetic structure: up to an overall constant, the components of $\Pi$ took values in the {\it cyclotomic field} ${\cal F}_d$, the extension of the rationals by $d^{th}$ roots of unity, with $d$ depending on the geometry. 

This field can be viewed as a vector space over the rationals; hence $f \cdot \Pi$ and $h \cdot \Pi$ can be expanded in a basis for this vector space, where the basis vectors of ${\cal F}_d$ are generally irrational numbers, and their rational coefficients are determined by $f$ and $h$. The existence of $W=0$ vacua is then conditional on whether integer fluxes $f$, $h$ can be found to cancel independently every basis vector in this space.  If this is done, one may always use the final $D_\psi W = 0$ equation to solve for the dilaton $\phi$, as $\phi$ enters in no other equations.  

The ability to find $W=0$ vacua thus depends on the dimension of this ``vector space of the periods", as a single equation like $f \cdot \Pi = 0$ will decompose into a number of relations on the fluxes equal to this dimension.  The hypersurface called $M_6$ in \cite{DGKT} had dimension two, and solutions were possible, while for the other three cases, the dimension was four and not enough freedom was present in the fluxes to find solutions.  Thus the four one-parameter models, naively quite similar, had dramatically different spectra of $W=0$ vacua.

\subsection{R-symmetries}

The $W=0$ vacua in the $M_6$ model are accompanied by a discrete R-symmetry,
a transformation of the moduli under which $W$ changes by a phase,
 ``enforcing" the vanishing of the superpotential at its fixed point $\psi = 0$.  It is worth explaining what constitutes a discrete symmetry in these models.  

The string theory compactifications we consider each possess a {\it modular group} of symmetries acting both on the moduli $z^a, \phi$ and on the fluxes $f$, $h$.  The $SL(2,\bZ)$ S-duality of type IIB string theory is one example, as are geometric transformations of the complex structure moduli space.  These symmetries are most usefully thought of as discrete {\it gauge} symmetries, since they correspond to a redundancy of the description: two apparent vacua related by such a transformation actually constitute only one genuine vacuum, and this redundancy must be accounted for to properly count vacua.

The symmetries we are interested in are not modular transformations, though they are related.  We want to consider {\it global} symmetries of the low-energy effective field theory for the moduli.  In this effective theory, the fluxes appear not as fields, but as coupling constants.  Consequently our R-symmetries will transform the moduli fields, but in order to stay within a given effective field theory description, they must not act on the couplings. 

In fact, the R-symmetries we find act on the moduli in the same fashion as the modular group, but leave out the transformation on the fluxes.  Unlike the modular transformations, which are always symmetries, such a moduli-only transformation need not be a good symmetry: they are symmetries only for special values of the fluxes.  When present, these are genuine global symmetries relating distinct vacua, not redunancies of the description.\footnote{Some models have modular transformations that happen to leave the fluxes invariant; these should be considered gauge symmetries as well, and represent an identification on the moduli space in the low-energy theory.}  Such a symmetry descending from a modular transformation can arise for values of the fluxes that place the vacuum on the fixed locus of that modular symmetry \cite{DGKT}, and we will use this princple in constructing $W=0$ vacua with R-symmetries.

\subsection{Extension to multiparameter cases}

The goal in this paper is to extend the construction of $W=0$ vacua with R-symmetries for one-parameter cases of \cite{DGKT} to hypersurfaces with additional complex structure parameters $\varphi^A$.  We shall continue to look at the Landau-Ginzburg point $\psi = 0$ and will look for R-symmetries descending from modular transformations rotating around this point.  The additional parameters complicate the analysis in two ways.  First, the periods now depend on the additional moduli $\varphi^A$, affecting the solution to the ``flux orthogonality" relations (\ref{TwoEqns}).  Second, additional equations $D_{\varphi^A} W = 0$ also need to be solved.

We shall describe how given a certain kind of solution to (\ref{TwoEqns})
the structure of the periods $\Pi$ are such that the $D_{\varphi^A} W = 0$ equations follow trivially.  The $\varphi^A$ moduli are then unconstrained by the fluxes, and the $D_\psi W = 0$ equation can be used to determine the dilaton as in the one-parameter case.  

Finding $W=0$ vacua thus will come down to finding these particular solutions of (\ref{TwoEqns}).  As we shall show, at $\varphi^A = 0$ the periods again take values in a cyclotomic field.  Away from the locus $\varphi^A = 0$, while they in general lack arithmetic properties, the periods still take values in a finite-dimensional vector space over the rationals.  The particular solutions in question are just choices of fluxes that solve (\ref{TwoEqns}) for this (in general larger) vector space at $\varphi^A \neq 0$.  As in the one-parameter case, where one out of four hypersurfaces supported $W=0$ vacua, we shall find that while not all spaces support such vacua, a substantial number do.

The construction we employ to obtain these vacua utilizes the matrix that generates monodromy around the $\psi = 0$ locus on moduli space, which will make manifest that all such $W=0$ vacua are accompanied by R-symmetries.  Before we can describe this construction, we must describe the geometries and their periods in more detail.

\section{Geometric considerations}
\label{GeomSec}

In this section we review relevant properties of hypersurfaces in weighted projective space, their mirrors and their periods.  Nothing in this section is new, though our emphasis on the ``period indices" $n_I$ and the Euler totient function $\phi(d)$ may be unfamiliar.

\subsection{Hypersurfaces in weighted projective space}

The period vector $\Pi(z)$ depends on the particular geometry considered.
The class of Calabi-Yau threefolds we work with are all hypersurfaces in weighted projective space.  The weighted projective space $\wcp^4_{k_1k_2k_3k_4k_5}$
has weighted homogeneous coordinates $x_i \sim \lambda^{k_i} x_i$ and degree $d \equiv \sum_{i=1}^5 k_i$.  We then define the Calabi-Yau manifold ${\cal M}$ as the vanishing locus (with singularities resolved) of the polynomial
\begin{eqnarray}
\label{Polynomial}
P(x_i) \equiv P_0(x_i) - d \psi x_1 x_2 x_3 x_4 x_5 + \sum_{\tilde{A}} \varphi^{\tilde{A}} M_{\tilde{A}}(x_i) \,.
\end{eqnarray}
Here $P_0(x_i)$ is a suitable ``defining polynomial"
\begin{eqnarray}
\label{DefinePoly}
P_0(x_i) = \sum_{j=1}^5 \prod_{i=1}^5 x_i^{a_{ij}} \,, \quad \quad 
\sum_{i=1}^5 k_i a_{ij} = d \quad \forall j \,.
\end{eqnarray}
The $M_{\tilde{A}}(x_i)$ are a set of monomials associated with complex structure variations $\varphi^{\tilde{A}}$:
\begin{eqnarray}
\label{othermonom}
M_{\tilde{A}}(x_i) = \prod_{i=1}^5 x_i^{q^i_{\tilde{A}}} \,, \quad \quad 
\sum_{i=1}^5 k_i q^i_{\tilde{A}} = d \quad \forall \,{\tilde{A}} \,,
\end{eqnarray}
and in (\ref{Polynomial}) we have separated out the  ``fundamental monomial" $M_0 =  -d \, x_1 x_2 x_3 x_4 x_5$ and denoted its modulus $\psi$.  In general not all complex structure deformations can be expressed in terms of a monomial $M_{\tilde{A}}$; these non-polynomial deformations are mirror to K\"ahler moduli that are not toric.  We shall restrict to the case of spaces with only monomial deformations here.

The simplest class of such hypersurfaces occurs when each $k_i$ divides $d$.  In this case, $a_{ij}$ can be chosen diagonal and 
$P_0$ is then a Fermat polynomial:
\begin{eqnarray}
P_0^{\rm Fermat}(x_i) = x_1^{d/k_1} + x_2^{d/k_2} + x_3^{d/k_3} + x_4^{d/k_4} + x_5^{d/k_5} \,.
\end{eqnarray}
We call these Fermat cases.  In the non-Fermat examples, all non-degenerate models have either a $0$ or a $1$ for the off-diagonal elements of $a_{ij}$, and take a block-diagonal form with Fermat blocks as well as either ``tadpole" $x_1^{a_{11}} x_2 + x_2^{a_{22}} x_3 + \ldots + x_n^{a_{nn}}$
or ``loop" $x_1^{a_{11}} x_2 + x_2^{a_{22}} x_3 + \ldots + x_n^{a_{nn}} x_1$
 blocks;  see for example \cite{Berglundetal}.

The defining polynomial $P_0$ (\ref{DefinePoly}) has a set of discrete phase rotation symmetries ${\cal T}_j$: each Fermat-type monomial $x_j^{a_{jj}}$ gives rise to a $\bZ_{a_{jj}}$, while tadpole and loop blocks give rise to $\bZ_{a_{11} a_{22} \ldots a_{nn}}$ and $\bZ_{a_{11} a_{22} \ldots a_{nn} + (-1)^{n-1}}$ symmetries, respectively. 
One linear combination of the ${\cal T}_j$, the ``quantum symmetry" $Q_{\cal M} \equiv \prod_j {\cal T}_j \simeq \bZ_d$, is a trivial identification of the homogeneous coordinates; the remaining symmetries of $P_0$ form the geometric symmetries $G_{\cal M}$.

The mirror of ${\cal M}$ is formed as follows\footnote{We assume $P$ is {\it transverse}; more general methods involving toric geometry are discussed for example in \cite{BKK}.} \cite{GreenePlesser, BerglundHubsch}.  Define the dual polynomial $\hat{P}_0$ with transposed exponents,  $\hat{a}_{ij} \equiv a_{ji}$:
\begin{eqnarray}
\label{MirrorDefinePoly}
\hat{P}_0(y_i) = \sum_{j=1}^5 \prod_{i=1}^5 y_i^{\hat{a}_{ij}} \,,
\end{eqnarray}
which lives in a projective space $\wcp^4_{\hat{k}_1\hat{k}_2\hat{k}_3\hat{k}_4\hat{k}_5}$ with coordinates $y_i \sim \lambda^{\hat{k}_i} y_i$ and degree $\hat{d} \equiv \sum_{i=1}^5 \hat{k}_i$.  The vanishing of this polynomial defines a space $\widehat{\cal W}$, and has phase symmetries $\widehat{\cal T}_j$ divided into $Q_{\widehat{\cal W}} = \bZ_{\hat{d}}$ and $G_{\widehat{\cal W}}$ as before.  The mirror ${\cal W}$ of ${\cal M}$ is then the quotient $\widehat{\cal W}/H$, where $H$ is the subgroup of $G_{\widehat{\cal W}}$ under which the fundamental monomal  $y_1 y_2 y_3 y_4 y_5$  is invariant.  One finds that $G_{\widehat{\cal W}} = H \times \bZ_d$, such that after the quotient one has the symmetries
\begin{eqnarray}
Q_{\cal W} = G_{\cal M} = H \times \bZ_{\hat{d}} \,, \quad G_{\cal W} = Q_{\cal M} = \bZ_d \,.
\end{eqnarray}
The polynomial $\hat{P}$ defining the mirror is then
\begin{eqnarray}
\label{MirrorPolynomial}
\hat{P} \equiv \hat{P}_0(y_i) - \hat{d} \psi y_0 y_1 y_2 y_3 y_4 + \sum_{\alpha = 1}^{m} \varphi^A M_A(y_i) \,,
\end{eqnarray}
where the $M_A(y_i)$ are monomials of degree $\hat{d}$ invariant under $H$.

In what follows we will work with the mirrors ${\cal W}$, for which there are a set of well-studied simple cases with a small number of complex structure moduli due to the restriction to $H$-invariant monomials; the corresponding ${\cal M}$ have a small number of K\"ahler moduli but in general many complex structure moduli, and are thus less useful as examples.  Our results should, however, generalize to the ${\cal M}$ models as well.

For Fermat cases, where the $ k_i$ divide $d$ and $a_{ij}$ is diagonal, we have $\hat{d} = d$ and ${\cal W}$ is just a quotient of ${\cal M} = \widehat{\cal W}$;  $\hat{P}$ is then simply a truncation of $P$ to monomials invariant under $H$, and one may use $x_i$ in place of $y_i$ since they are defined in the same space.  We may then think of merely restricting to $H$-invariant monomials in ${\cal M}$ instead of actually taking the mirror; as discussed in \cite{GKTT}, turning on $H$-invariant fluxes consistently leads to solutions in this subspace of the moduli space of ${\cal M}$.

\subsection{Periods}

The periods for ${\cal W}$ may then be defined as follows \cite{Candelasetal, Berglundetal, BK}.  One begins with the fundamental period $\varpi_0$, defined by a canonical choice of holomorphic three-form and of a $B_0$-cycle.  The integral over the cycle is evaluated and one obtains a series in $1/\psi$.  For $\varphi^A = 0$ this can be written as a generalized hypergeometric function:
\begin{eqnarray}
\label{hyperg}
\varpi_0 = \; _qF_{q-1}(\hat{n}_I ; \hat{m}_J, 1, 1, 1; (\prod_{i=1}^5 k_i^{k_i} \psi)^{-d} )\,,
\end{eqnarray}
where the $\{ \hat{n}_I\}$ and $\{\hat{m}_J\}$ are determined as follows: compare the set $\{ 1/d, 2/d, \ldots (d-1)/d \}$ with the set $\{ 1/k_1, \ldots (k_1-1)/k_1, \ldots 1/k_5, \ldots (k_5-1)/k_5\}$, and remove from both sets any number appearing in both (but only remove it once in each for each match); what remains are the $\hat{n}_I$ and the $\hat{m}_J$, respectively.  The integer $q \leq d-1$ is the number of the $ \hat{n}_I$.  Note that (\ref{hyperg}) is a period of the {\it mirror} ${\cal W}$, but is given in terms of the weights $k_i$ and degree $d$ of the original space ${\cal M}$.

The fundamental period can be analytically continued to the region of small $\psi$ via an integral of Barnes type; this is done in general in \cite{BK}.  The result is
\begin{eqnarray}
\label{fundperiodone}
\varpi_0 = - \sum_{n=1}^\infty {\Gamma({n \over d}) \alpha^{n(d-1)/2} (k_1^{k_1/d} d \psi)^{n-1} \over \Gamma(n) \prod_{r=1}^{k_1 -1} \Gamma({r \over k_1} - {n \over d}) \prod_{i=2}^5 \Gamma(1- {k_i n \over d}) } \,,
\end{eqnarray}
where $k_1$ is the smallest of the $k_i$, $\alpha$ is a $d^{th}$ root of unity:
\begin{eqnarray}
\label{alpharoot}
\alpha^d = 1 \,,
\end{eqnarray}
and we have renormalized by an extra factor $(2 \pi)^{(1-k_1)/2} /\psi$ relative to (\ref{hyperg}) and to \cite{BK}; this K\"ahler transformation assures the periods and thus the K\"ahler potential (\ref{KahlerPot}) are regular at $\psi =0$, as $W=0$ is not predictive if $e^K$ diverges.

In what follows we specialize for simplicity to $k_1 = 1$, which is obeyed by all our examples.  The inclusion of the neglected parameters $\varphi^A$ then gives the fundamental period
\begin{eqnarray}
\label{fundperiod}
\varpi_0 = - \sum_{n=1}^\infty {\Gamma({n \over d}) \alpha^{n(d-1)/2} (d \psi)^{n-1} \over \Gamma(n) \prod_{i=2}^5 \Gamma(1- {k_i n \over d}) } U_n(\varphi^A) \equiv \sum_{n=1}^\infty c_n U_n(\varphi^A) \psi^{n-1} \,,
\end{eqnarray}
with $U_n(\varphi^A = 0) = 1$; this will in general only be well-defined in some neighborhood of $\varphi^A = 0$.  The functional form of the $U_n(\varphi^A)$ will not matter for us.

The remaining periods are generated from the fundamental period $\varpi_0$ by a monodromy action associated to the discrete symmetries $G_{\cal W} \simeq \bZ_d$.  The full polynomial (\ref{Polynomial}) with nonzero $\psi, \varphi^A$ is not preserved by $G_{\cal W}$ transformations; instead, the $G_{\cal W}$ action can be canceled by also transforming the moduli $\psi, \varphi^A$.  This generates an identification on the moduli space of $\psi, \varphi^A$, since transforming the moduli can be absorbed into a change of coordinates.  This is a subgroup ${\cal A}$ of the full monodromy group of the moduli space, with ${\cal A} \simeq G_{\cal W} \simeq \bZ_d$.

The monomials $M_A$ invariant under $H$ transform nontrivially under $G_{\cal W}$.  In particular, the fundamental monomial always faithfully represents $\bZ_d$.  This induces the monodromy action on the moduli space:
\begin{eqnarray}
\label{monodromy}
{\cal A}: \quad \psi \to \alpha \psi \,, \quad \varphi^A \to \alpha^{Q_A} \varphi^A \,.
\end{eqnarray}
The periodicities of the $\varphi^A$ determined by $Q_A$ will be important for us.  For cases with $k_1 = 1$ one can take $\bZ_d$ to be generated by $x_1 \to \alpha^{-1} x_1$, in which case we have $Q_A = q_A^1$, with $q_A^1$ as in (\ref{othermonom}).

From the fundamental period one may then use the monodromy ${\cal A}$ to define $d-1$ additional periods:
\begin{eqnarray}
\label{otherperiods}
\varpi_J(\psi, \varphi^A) \equiv \varpi_0(\alpha^J \psi, \alpha^{Q_A J} \varphi^A) = \sum_{n=1}^\infty c_n \alpha^{n J} U_n(\alpha^{Q_A J} \varphi^A)  \psi^{n-1}  \,.
\end{eqnarray}
In general not all $d$ of the periods so constructed are independent, instead $q$ are independent where $q$ is the index of the hypergeometric (\ref{hyperg}) \cite{Berglundetal}.  This generates all the periods associated to complex structure moduli that can be written as monomial deformations of the fundamental polynomial; since we focus on examples where all moduli are of this sort, we have generated all periods from (\ref{otherperiods}) and write $b_3 = q$.
We can choose a basis $\varpi_0, \ldots \varpi_{b_3-1}$ and arrange them into a vector $\vec\varpi$:
\begin{eqnarray}
\label{periodvec}
\vec\varpi(\psi, \varphi^A) = \sum_{n=1}^\infty c_n \vec{p}_n (\varphi^A) \psi^{n-1} \,, \quad \quad
c_n = -
{\Gamma({n \over d}) \alpha^{n(d-1)/2} d^{n-1} \over \Gamma(n) \prod_{i=2}^5 \Gamma(1- {k_i n \over d}) } 
\,.
\end{eqnarray}
with $(p_n)_J (\varphi^A) \equiv \alpha^{n J} U_n(\alpha^{Q_A J} \varphi^A)$.  We shall have special use for the term at $\psi = 0$, and denote the corresponding functions at $n=1$ as $p_J(\varphi^A) = \alpha^J U(\alpha^{Q_A J} \varphi^A)$.

Finally, the symplectic basis for the periods $\Pi$ (which enters into the equations (\ref{DWZero}, \ref{WZero}) for vacua) is obtained from the Picard-Fuchs basis $\varpi$ via a transformation
\begin{eqnarray}
\label{PiBasis}
\Pi_I(\psi, \varphi^A) = m_{IJ} \varpi_J(\psi, \varphi^A) \,,
\end{eqnarray}
where the matrix $m_{IJ}$ has {\it rational} entries \cite{Morrison}.  As we shall discuss, the precise form of $m_{IJ}$ will not be important for us.  The expression (\ref{periodvec}), (\ref{PiBasis}) for the periods is the result we will use in the following sections.

Note that not all values of $n$ in the expansion (\ref{periodvec}) have nonzero coefficients $c_n$.  The $\Gamma$-functions in the denominator have poles, killing the coefficient, whenever
\begin{eqnarray}
\label{poles}
n = {d \ell \over k_i} \,,
\end{eqnarray}
for any positive integer $\ell$ and any $k_i$ with $i = 2,3,4,5$.  It is easy to see that the values of $n$ having nonzero $c_n$ are periodic mod $d$; and moreover those values $1 \leq n \leq d$ with nonzero $c_n$, call them the $n_I$, are precisely the integers
\begin{eqnarray}
\label{nonzero}
n_I = d \hat{n}_I \,,
\end{eqnarray}
with the $\hat{n}_I$ as given below (\ref{hyperg}).  We call the $n_I$, which are $b_3$ in number, the {\it period indices}.

The period indices will play an essential role in our analysis, so we mention a few of their properties.  They are determined solely by the $k_i$ defining the weighted projective space.  The set contains at minimum those integers $1 \leq n_I < d$ that share no common factors with $d$.  The number of such integers is the definition of the {\it Euler totient function} $\phi(d)$. Hence we have in general
\begin{eqnarray}
\phi(d) \leq b_3 < d \,.
\end{eqnarray}
As we shall see, whether $b_3 > \phi(d)$ or $b_3 = \phi(d)$  will determine whether there may be or may not be R-symmetric $W=0$ vacua at $\psi = 0$ in a given model.

\section{Construction of $W=0$ vacua}
\label{ConstructSec}

With this description of the periods in hand, we go on to construct vacua satisfying $DW = W = 0$.  The equations $D_\phi W = W = 0$ (\ref{TwoEqns}) at $\psi = 0$ become using (\ref{periodvec})
\begin{eqnarray}
\label{SatisfyTwoEqns}
\tilde{f} \cdot p (\varphi^A) = \tilde{h} \cdot p(\varphi^A) = 0 \,,
\end{eqnarray}
where $p(\varphi^A)$ is (up to an overall coefficient) the $\varpi$ period vector at $\psi =0$, and $\tilde{f}_J \equiv f_I \, m_{IJ}$ and $\tilde{h}_J \equiv h_I \, m_{IJ}$ are vectors of rational numbers.  
If a solution to these ``flux orthogonality" equations exists for rational $\tilde{f}$, $\tilde{h}$, we can always scale up to find a set of integer solutions for $f$, $h$; hence merely to demonstrate existence of vacua, we do not need to know the values of $m_{IJ}$, just that they are rational. We first discuss the algebraic properties of these equations, before turning to the remaining $DW=0$ equations and a construction for a solution.

\subsection{The vector space of periods}

For given $\varphi^A$, the elements of the vector $p_J(\varphi^A) \equiv \alpha^J U(\alpha^{Q_A J} \varphi^A)$ are in general not rational, but instead generate a vector space ${\cal V}$ over the rationals, the ``vector space of periods", with basis
\begin{eqnarray}
\label{VBasis}
{\rm basis}\ {\cal V} = 
\{ U(\varphi^A), \alpha U(\alpha^{Q_A} \varphi^A), \ldots \alpha^{b_3 - 1} U(\alpha^{Q_A (b_3 -1)} \varphi^A) \} \,.
\end{eqnarray}
Satisfying (\ref{SatisfyTwoEqns}) by tuning the rational rescaled fluxes $\tilde{f}$, $\tilde{h}$ can only be done by separately setting to zero the rational coefficient of each basis element of this vector space.  It is thus vital for the possibility of a solution that the $p_J$ (\ref{VBasis}) are not linearly independent over the rationals; otherwise the only solution would be that all fluxes vanish.   This lack of independence will arise because of a combination of two effects.

First, the $d$ basis elements of the cyclotomic field ${\cal F}_d$, $\{ 1, \alpha, \alpha^2, \ldots \alpha^{d-1} \}$, are in general not independent when $d$ is not prime; only $\phi(d)$ are independent, where $\phi(d)$ is the Euler totient function defined in the last section.  In one-parameter models, the functions $U$ are absent and the vector space of periods ${\cal V}$ is simply the cyclotomic field ${\cal F}_d$ extending the rationals by the root of unity $\alpha$ \cite{DGKT}, and one has $\dim {\cal V} = \phi(d)$.

Secondly, in the multiparameter cases one also has the functions $U(\varphi^A)$, which in general take transcendental values.  
However, the $U(\alpha^{Q_A J}\varphi^A)$ need not be distinct for all $J$, as $Q_A$ may share a common factor with $d$, leading to $U(\alpha^{Q_A J}\varphi^A) = U(\alpha^{Q_A (J+D_A)}\varphi^A)$ for some $D_A$.  As a result, the function $U(\varphi^A)$ has an overall {\it periodicity} set by $D = \lcm(\{D_A\})$.  Thus we find that $p_{J+D}(\varphi^A) = \alpha^D p_J(\varphi^A)$, and such elements may be linearly dependent regardless of the value of the $\varphi^A$.

Hence the $p_J(\varphi^A)$ will in general obey linear relationships for {\it arbitrary values of} $\varphi^A$, defining a vector space ${\cal V}$ over the rationals with $\varphi^A$-dependent basis elements, satisfying
\begin{eqnarray}
\label{Inequality}
\phi(d) \leq \dim {\cal V} \leq b_3 < d \,.
\end{eqnarray}
By tuning the $f$ and $h$ fluxes to obtain a zero coefficient for each basis element of ${\cal V}$, one finds a solution for (\ref{SatisfyTwoEqns}) for any $\varphi^A$.  If the dimension of ${\cal V}$ is equal to $b_3$, this is only possible for $f = h = 0$.  On general grounds one then expects that the dimension of the space of orthogonal flux vectors  solving either equation in (\ref{SatisfyTwoEqns}) is given by
\begin{eqnarray}
\label{NumSolns}
b_3 - \dim {\cal V} \,. 
\end{eqnarray}
For fixed $d$ and $b_3$ this is largest when $\dim {\cal V} = \phi(d)$, which can only occur when the overall periodicity $D$ is a factor of $\phi(d)$.

As an example, consider the two-parameter model $(k_1, k_2, k_3, k_4, k_5) = (1,1,2,2,6)$; this was studied in \cite{GKTT} and will be further analyzed in section \ref{ExampleSec}. This model has $d=12$, and the cyclotomic field ${\cal F}_{12}$ has $\phi(12) = 4$ linearly independent elements, which can be taken to be $1, \alpha, \alpha^2, \alpha^3$ with the rest related via
\begin{eqnarray}
\label{CycloTwelve}
\alpha^4 = \alpha^2 - 1 \,, \quad \quad {\rm for}\ \alpha^{12} = 1 \,.
\end{eqnarray}
The monomial associated to the parameter $\varphi$ is $M = x_1^6 x_2^6$ with $Q = q^1 = 6$, meaning $U(  \alpha^{Q} \varphi) = U(-\varphi)$, hence with overall periodicity $D=2$.  For a two-parameter model there are $b_3 = 6$ elements of $p_J(\varphi)$ generating the vector space of periods ${\cal V}$,
\begin{eqnarray}
{\rm basis}\ {\cal V}_{11126} = \{ U(\varphi), \alpha U(-\varphi), 
\alpha^2 U(\varphi), \alpha^3 U(-\varphi), 
\alpha^4 U(\varphi), \alpha^5 U(-\varphi) \} \,,
\end{eqnarray}
but using (\ref{CycloTwelve}), only the first four are independent; hence in this case $\dim {\cal V}  = \phi(d) = 4$, thanks to the compatibility between the overall periodicity $D$ and the dimension of the cyclotomic field $\phi(d)$.

For a more complicated case where $\dim {\cal V} > \phi(d)$, consider another $d=12$ space, $(1,2,3,3,3)$.  The cyclotomic field is the same as the previous example, and the relation (\ref{CycloTwelve}) still holds, but this time there are two additional parameters $\varphi^1$, $\varphi^2$ with monomials $M_1 = x_1^4 x_2^4$ and $M_2 = x_1^8 x_2^2$, with $Q_1 = 4$ and $Q_2 = 8$, giving rise to overall periodicity $D = 3$.  This three-parameter model has the vector space of periods generated by the $b_3 = 8$ elements of $p_J(\varphi^A)$,
\begin{eqnarray}
{\rm basis}\ {\cal V}_{12333} &=& \{ U(\varphi^1, \varphi^2), \alpha U(\alpha^4 \varphi^1, \alpha^8 \varphi^2), 
\alpha^2 U(\alpha^8 \varphi^1, \alpha^4 \varphi^2), \alpha^3 U(\varphi^1, \varphi^2), \\ && \hskip-.5in
\alpha^4 U(\alpha^4 \varphi^1, \alpha^8 \varphi^2), \alpha^5 U(\alpha^8 \varphi^1, \alpha^4 \varphi^2),
- U(\varphi^1, \varphi^2),- \alpha U(\alpha^4 \varphi^1, \alpha^8 \varphi^2)
 \} \,,\nonumber
\end{eqnarray}
where thanks to $\alpha^6 = -1$ we see the last two elements are not independent; however despite the relation (\ref{CycloTwelve}) we cannot write the $\alpha^4$ and $\alpha^5$ elements in terms of the others, because the periodicity of the $U$'s is not compatible, a reflection of $\phi(d) = 4$ not being a multiple of $D = 3$.  Hence in this example, $\dim {\cal V} = 6 > \phi(d) = 4$.

At a specific value of $\varphi^A$ the vector space in which the periods live may become smaller; for example at $\varphi^A = 0$ it reduces simply to ${\cal F}_d$.  
A more complicated example is described for the $(1,1,2,2,6)$ two-parameter model in sec.~8.3 of \cite{AA}, where it is shown that solving the conditions for ``attractor points" at $\psi = 0$ leads to a set of discrete values for $\varphi$ where the periods take values in the rank two vector space (and field) $\bQ[i]$.  However, as we now discuss, solving the additional $D_{\varphi^A} W = 0$ equations in the fashion we describe necessitates a solution valid for a continuous range of  $\varphi^A$.

\subsection{Solving remaining complex structure equations}

We will find that solutions to the flux orthogonality equations (\ref{SatisfyTwoEqns}) for arbitrary $\varphi^A$ are not uncommon.  Consider such a solution, and try to solve the equations $D_{\varphi^A} W = 0$. Then since $W=0$, each equation reduces to
\begin{eqnarray}
\label{PartialEqns}
\partial_A W = (\tilde{f} - \phi \tilde{h}) \cdot \partial_A p(\varphi^A) = 0\,.
\end{eqnarray}
One has $\partial_A p_J(\varphi^A) = \alpha^J \partial_A U(\alpha^{Q_A J} \varphi^A)$.  The essential point is that for each $A$, the $\partial_A p_J(\varphi^A)$ will generate a vector space ${\cal V}^{(A)}$
with basis elements obeying the {\it same} linear relationships as the basis elements of ${\cal V}$ generated by the $p_J(\varphi^A)$. 

This occurs because the $\alpha^J$ coefficient is does not change, and the periodicities of the $U$'s -- which are all that go into determining the linear relations on the basis vectors -- are not disturbed by the derivative.  So although the elements of the basis of the space ${\cal V}^{(A)}$ will in general be different transcendental numbers than those of ${\cal V}$, their coefficients in (\ref{PartialEqns}) will be zero if the coefficients in (\ref{SatisfyTwoEqns}) are.

Moreover, the $f$ and $h$ parts are separately zero, meaning that again the dilaton $\phi$ drops out of the equation.  Since the dilaton is completely unconstrained thus far, we may simply use the final equation $D_\psi W = 0$ to solve for it:
\begin{eqnarray}
\label{DilatonEqn}
\phi = {f \cdot \partial_\psi \Pi \over h \cdot \partial_\psi \Pi} \,.
\end{eqnarray}
Hence we find that once a solution to (\ref{SatisfyTwoEqns}) with the periods orthogonal to the flux vectors is found for arbitrary $\varphi^A$ , the remaining $DW=0$ equations can always be satisfied.   

One can consider finding a solution to (\ref{SatisfyTwoEqns}) valid only at a single point $\varphi^A_0$; however if this solution does not extend to a continuous family of solutions over the space of $\varphi^A$, the $D_{\varphi^A}  W = 0$ solutions will not in general be satisfied.  It is not impossible that the ranks of ${\cal V}$ and of the ${\cal V}^{(A)}$ may all reduce at the same point, or ${\cal V}$ and the ${\cal V}^{(A)}$ may at a certain point become isomorphic.  Such circumstances would lead to isolated $W=0$ vacua; whether they exist is an interesting open questions.

\subsection{Ansatz for vacua}

Consequently, finding $W=0$ vacua reduces to solving the flux orthogonality equations (\ref{SatisfyTwoEqns}) over a continuous range of $\varphi^A$.
To do so, we are interested finding integer $b_3$-vectors $g_J$ obeying
\begin{eqnarray}
\label{VPi}
g \cdot \Pi(\psi = 0, \varphi^A) = 0 \,.
\end{eqnarray}
A solution to (\ref{SatisfyTwoEqns}) will then follow setting $f_J = g_J^{(1)}$  and $h_J = g_J^{(2)}$ for two such vectors $g^{(1)}$ and $g^{(2)}$, which must be non-parallel in order to produce nonzero tadpole (\ref{Tadpole}).  
Our method for finding such a vector is essentially to require the existence of an R-symmetry in the low-energy effective theory.  We do this as follows.

Consider the monodromy action ${\cal A}$ (\ref{monodromy}) generating phase rotations around the LG point $\psi = 0$.  It will be realized on the periods $\Pi_J$ as a $b_3 \times b_3$ matrix $A_{IJ}$ representing $\bZ_d$, hence obeying $A^d = 1$.  It has the general form
\begin{eqnarray}
\label{AMatrix}
A =  m \cdot \begin{pmatrix}
\;0\;\;\; 1\;\;\; 0 \;\;\; \cdots  \;\;\; 0 \; \cr 
\;0\;\;\; 0\;\;\; 1 \;\;\; \cdots \;\;\; 0 \; \cr
 \vdots \;\;\; \;\vdots \;\;\;\; \vdots \;\;\;\; \cdots \;\;\; \vdots \; \cr
\;0\;\;\; 0\;\;\; 0 \;\;\; \cdots \;\;\; 1 \; \cr
\;*\;\; *\;\;* \;\;\; \cdots \;\;\; * \; 
\end{pmatrix} \cdot m^{-1} \,,
\end{eqnarray}
where the rational entries on the final line are determined by the expansion of $\varpi_{b_3}$ in the other $\varpi_J$ for each particular model.  The 
monodromy action of this matrix on the periods $\Pi$ translates into 
\begin{eqnarray}
\label{AAction}
A \Pi(\psi, \varphi^A) = \alpha \Pi(\alpha \psi,\alpha^{Q_A} \varphi^A) \,,
\end{eqnarray}
where the overall factor of $\alpha$ comes from our normalization of the $\varpi_J$. 

As described in section \ref{VacuaSec}, an R-symmetry must be a transformation on the moduli that leaves the fluxes invariant, such that the superpotential rotates by a nontrivial phase.  Consider the action of ${\cal A}$ (\ref{monodromy}) some number $N$ times on the moduli, while holding the fluxes invariant.  One has
\begin{eqnarray}
\label{WTrans}
W(\phi, \psi, \varphi^A) = (f - \phi h) \cdot \Pi(\psi,\varphi^A) \to \alpha^{-N} \, (f - \phi h) \cdot A^N \cdot \Pi(\psi,\varphi^A) \,.
\end{eqnarray}
This can become an overall phase rotation of $W$ if $f$ and $h$ are left eigenvectors of $A^N$.

As $A^d = 1$, the eigenvalues of $A$ and its powers are $d^{th}$ roots of unity.
Since $A$ and $f$, $h$ all contain rational numbers, the only possible eigenvalues are $\pm 1$.  Correspondingly, our ansatz consists of looking for flux vectors $g$ satisfying the ``R-symmetry constraint"
\begin{eqnarray}
\label{VEqn}
g = g \cdot A^N \,,
\end{eqnarray}
for some integer $1 \leq N < d$.
An R-symmetry implies the vanishing of the superpotential when the vacuum lies on the fixed locus of the symmetry, and $g \cdot \Pi$ vanishes for analogous reasons. Given (\ref{AAction}) and (\ref{VEqn}), we have 
\begin{eqnarray}
\label{EvalV}
g\cdot \Pi(\psi =0, \varphi^A) = g \cdot A^N \cdot \Pi(\psi=0,\varphi^A) = \alpha^N g \cdot \Pi(\psi = 0, \alpha^{N Q_A} \varphi^A) \,.
\end{eqnarray}
We then find that flux orthogonality (\ref{VPi}) follows if the R-symmetry constraint (\ref{VEqn}) is accompanied by an additional {\it periodicity condition}
\begin{eqnarray}
\label{PeriodCond}
\alpha^{N Q_A} = 1  \quad\quad  \forall \, Q_ A \,.
\end{eqnarray}
As discussed, the $\varphi^A$ have an overall periodicity $D$; (\ref{PeriodCond}) then requires $N = \ell D$ for integer $\ell$.
Taking $f$, $h$ proportional to such $g$'s then implies the existence of a $\bZ_{d/N}$ R-symmetry
\begin{eqnarray}
\label{RSym}
\psi \to \alpha^N \psi \,, \quad \varphi^A \to \varphi^A  \quad \Rightarrow \quad W \to \alpha^{-N} \, W \,.
\end{eqnarray}
Since powers of $A$ may also have eigenvalue $-1$, it is possible that the square root of the R-symmetry constraint may also hold:
\begin{eqnarray}
g = - g \cdot A^{N/2} \,.
\end{eqnarray} 
Hence if $N/2$ is even, the R-symmetry may be promoted to $\bZ_{2 d/N}$:
\begin{eqnarray}
\label{RSymTwo}
\psi \to \alpha^{N/2} \psi \,, \quad \varphi^A \to \alpha^{N Q_A/2} \varphi^A  \quad \Rightarrow \quad W \to -\alpha^{-N/2} \, W \,.
\end{eqnarray}
Note that $N/2$ need not necessarily be compatible with the periodicity condition.

\subsection{The R-symmetry constraint and eigenvalues of $A$}

Obtaining $W=0$ vacua thus comes down to imposing the R-symmetry constraint (\ref{VEqn}) while satisfying the periodicity condition (\ref{PeriodCond}).
To understand when we can impose the R-symmetry constraint, we consider the eigenvalues of $A$.  

Since $A^d = 1$, each eigenvalue is a $d^{th}$ root of unity, $\alpha^l$ for some $l$.  In principle we can calculate these by determining the relation between $\varpi_{b_3}$ and the other $\varpi_J$ to fill in the last line of the matrix (\ref{AMatrix}), and diagonalize it.
However, the eigenvalues may be determined more simply either by using the explicit formula for the periods $\varpi$ (\ref{otherperiods}) expanded in a power series in $\psi$, or simply from the monodromy action (\ref{AAction}).  Using these one can show that
\begin{eqnarray}
\Pi^{(n)} \equiv  \sum_{J=0}^{b_3 -1} \partial_\psi^n \Pi(\psi, \alpha^{Q_A J} \varphi^A) \Big|_{\psi = 0} \propto c_{n+1} \, \sum_{J=0}^{b_3-1} p_{n+1}(\alpha^{Q_A J} \varphi^A) \,,
\end{eqnarray}
is an eigenvector of $A$, with eigenvalue $\alpha^{n+1}$.  

However, not all $\Pi^{(n)}$ are nonzero as some of the coefficients $c_n$ vanish.  The coefficients are nonzero precisely for the values $n$ mod $d = n_J$, with the period indices $n_J$ given in (\ref{nonzero}).  Hence the eigenvalues $a_J$ of $A$ are just 
\begin{eqnarray}
 a_J  =  \alpha^{n_J} \,.
 \end{eqnarray}  
We have established previously that there are exactly $b_3$ nonzero $n_J$, giving the correct number of eigenvalues for $A$.

These eigenvalues represent $\bZ_d$ and hence always obey $a_J^d = 1$.  However, it is not necessarily the case that a given eigenvalue {\it faithfully} represents $\bZ_d$.  When $d$ is not prime, there exist powers $\alpha^l$ of $\alpha$ with $l$ sharing a factor with $d$; one then has $(\alpha^l)^{d/m} = 1$ where $m = \gcd(l, d)$, and the eigenvalue represents $\bZ_{d/m}$.

Eigenvalues that faithfully represent $\bZ_d$ will not help in imposing (\ref{VEqn}),
as there is no power $N < d$ for which they become unity.  Instead,
unfaithful eigenvalues are what we are looking for: by definition there exists a power 
\begin{eqnarray}
N_J \equiv { d \over \gcd(a_J, d)}  < d
\end{eqnarray}
such that $a_J^N = +1$.
Thus we find the mechanism we need: for each unfaithful eigenvalue $a_J$ of $A$, there exists an $N_J$ such that $A^{N_J}$ has one nontrivial eigenvector $g$ and the R-symmetry constraint (\ref{VEqn}) can be imposed.  Let us try to count these unfaithful eigenvalues.

As remarked previously, the set $\{ n_J \}$ of period indices always contains at least the integers $1 \leq n_J < d$ that share no common factors with $d$.  These are precisely the powers leading to faithful eigenvalues $\alpha^{n_J}$: hence the (unhelpful) faithful eigenvalues are always present.  Since the number of faithful $n_J$ is the definition of the Euler totient function $\phi(d)$, and there are $b_3$ of the $n_J$, we have
\begin{eqnarray}
\phi(d) \leq b_3 < d \,.
\end{eqnarray}
The remaining $b_3- \phi(d)$ eigenvalues never faithfully represent $\bZ_d$, since the exponents share a common factor with $d$. Hence, before taking into account the
periodicity condition (\ref{PeriodCond}), we expect $b_3 - \phi(d)$ independent solutions to (\ref{VEqn}).

In fact, one may quickly determine whether there are unfaithful eigenvalues of the monodromy matrix simply by looking at the weights $k_i$, as follows.  Consider the prime factors $p_\alpha$ of the degree $d$.  If any of the integers $d/p_\alpha$ is {\it absent} from the set $\{ k_i \}$, there is an unfaithful eigenvalue $\alpha^{p_\alpha}$ (and in general  others of the form $\alpha^{m p_\alpha}$ for some $m$) that can be used to impose the R-symmetry constraint (\ref{VEqn}) with $N = d/p_\alpha$.   If all integers $d/p_\alpha$ are found among the $k_i$, all eigenvalues are faithful.

The conclusion of the study of the one-parameter cases \cite{DGKT} was that the dimension of the space of fluxes solving $DW = W = 0$ was given by the difference of the number of fluxes, and the number of constraints placed on them by expanding the equations in the basis of the extension over the rationals in which the periods lived -- there just the cyclotomic field ${\cal F}_d$.  For each $f$ and $h$ flux, this difference was just $b_3 - \phi(d)$, consistent with the results found here.

In the multiparameter case, however, one must also confront the periodicity condition (\ref{PeriodCond}), as we now discuss.

\subsection{The periodicity condition}

For Fermat models, the periodicity $D$ of the moduli $\varphi^A$ can be calculated from the $k_i$ as follows.  
Consider all sets of 2 or 3 of the $k_i$ and find the greatest common divisor.  When this is not unity, call it $D_A$.  There is then a corresponding monomal $M_A$, composed only of the coordinates $x_i$ whose indices do {\it not} contain $D_A$, of the form (\ref{othermonom}) with
\begin{eqnarray}
q^i_A = {d \over D_A} + {d \, b^i_A \over \lcm(D_A, \, k_i) } \,,
\end{eqnarray}
for any integers $b^i_A > - \lcm(D_A, k_i)/D_A$, with total degree $d$ if $\sum_i (k_i + D_A k_i b^i_A / \lcm(D_A, k_i) ) = D_A$. (In general there can be more than one admissible monomial if several values of $b^i_A$ are permitted.)  It is then easy to see that $D_A$ is the periodicity of this monomial, and hence the modulus $\varphi^A$ as well.
Consequently the total periodicity is simply\footnote{For Fermat examples this coincides with the weight of the hyperplane class of ${\cal M}$ \cite{BK}.}
\begin{eqnarray}
D \equiv \lcm(\{ D_A \}) \,,
\end{eqnarray}
and the periodicity requirement for an eigenvalue $\alpha^{n_J}$ is that $N_J = \ell_J D$ for some integer $\ell_J$.

Hence for Fermat cases, one may deduce whether there are $W=0$ vacua entirely from the $k_i$.  As discussed in the last subsection, any integer $d/p_\alpha$ with $p_\alpha$ a prime factor of $d$ that is absent from the $k_i$ becomes a value of $N$ for a set of unfaithful eigenvalues; the periodicity condition is satisfied if this $d/p_\alpha$ is a multiple of all common factors of the set of $k_i$.
For non-Fermat cases, the monomials live in $\wcp^4_{\hat{k}_1,\hat{k}_2,\hat{k}_3,\hat{k}_4,\hat{k}_5}$ and cannot be analyzed so simply; we comment on a few non-Fermat examples in the next section.

The requirement of the periodicity condition may remove dimensions from the space of possible fluxes.  It corresponds precisely to the fact that the vector space of periods ${\cal V}$ can have dimension larger than the cyclotomic field ${\cal F}_d$, and hence can place additional constraints on the fluxes.  Instead of $b_3 - \phi(d)$, the more general formula is $b_3 - \dim {\cal V}$, corresponding to the number of eigenvalues that unfaithfully represent $\bZ_d$ as well as complying with the periodicity condition (\ref{PeriodCond}) on $U(\varphi^A)$. 
This is our result for the incidence of R-symmetric $W=0$ vacua in hypersurfaces in weighted projective space, whose agreement with the 
 counting (\ref{NumSolns}) indicates that the construction via the R-symmetry constraint (\ref{VEqn}) produces all the relevant vacua.

\subsection{Enhanced symmetries for special values of the dilaton}

The R-symmetries we have constructed all involve a transformation of the complex structure moduli, with the dilaton inert.  Imposing additional constraints on the fluxes, one may enhance these symmetries by allowing the dilaton to transform.

These transformations are associated with vacua where the dilaton sits at one of the fixed points of the $SL(2,\bZ)$ modular group, either the $\bZ_2$ fixed point $\phi = i$ or the $\bZ_3$ fixed point $\phi = \exp(\pi i /3)$.  Consider a transformation including a $\bZ_2$ action on the dilaton,
\begin{eqnarray}
\label{ZTwoTrans}
\psi \to \alpha^M \psi \,, \quad \varphi^A = \alpha^{Q_A M} \varphi^A \,, \quad \phi \to - {1 \over \phi} \,.
\end{eqnarray}
If the fluxes satisfy the constraint
\begin{eqnarray}
\label{ZTwoConstraint}
h \cdot A^M = \pm f \,, \quad \quad f \cdot A^M = \mp h \,,
\end{eqnarray}
the transformation of the superpotential becomes
\begin{eqnarray}
W(\psi, \varphi^A, \phi) \to  \pm {\alpha^{-M} \over \phi} W(\psi, \varphi^A, \phi)\,.
\end{eqnarray}
Because the dilaton transformation acts as K\"ahler transformation on the K\"ahler potential, the superpotential must transform with the factor $1/\phi$ for the total to be a symmetry.

The constraints (\ref{ZTwoConstraint}) require that $A^{2M}$ has eigenvalue $-1$.  Hence we see that if an R-symmetry associated to a relation $g = g \cdot A^N$ obeys $N = 4M$ for integer $M$, we can extend the $\bZ_{d/N}$ R-symmetry to 
its ``fourth root", the larger $\bZ_{d/M} = \bZ_{4d/N}$ transformation (\ref{ZTwoTrans}). 

Using the flux constraints (\ref{ZTwoConstraint}) and the formula (\ref{DilatonEqn}) for the dilaton, one may show that $\phi = i$ in the vacuum; hence as with the complex structure moduli space, the symmetries are present at fixed points of the modular group action.

An analogous situation holds for the $\bZ_3$ transformation on the dilaton
\begin{eqnarray}
\label{ZThreeTrans}
\psi \to \alpha^M \psi \,, \quad \varphi^A = \alpha^{Q_A M} \varphi^A \,, \quad \phi \to {1 \over 1- \phi} \,.
\end{eqnarray}
In this case the flux constraint is
\begin{eqnarray}
\label{ZThreeConstraint}
h \cdot A^M = \pm (h-f) \,, \quad \quad f \cdot A^M = \pm h \,,
\end{eqnarray}
and the transformation of the superpotential becomes
\begin{eqnarray}
W(\psi, \varphi^A, \phi) \to  \pm {\alpha^{-M} \over 1- \phi} W(\psi, \varphi^A, \phi)\,.
\end{eqnarray}
In this case the requirement is that $A^{3M}$ has eigenvalue $\mp 1$.  Hence for an R-symmetry $\bZ_{d/N}$ associated to $g =  g \cdot A^N$ with $N=3M$, one can promote it to a symmetry $\bZ_{3d/N}$ (\ref{ZThreeTrans}) that cubes to the original R-symmetry.  Similarly, a symmetry $\bZ_{2 d/N}$ corresponding to $g = - g \cdot A^{N/2}$ can be promoted to $\bZ_{6d/N}$.
The flux constraints (\ref{ZThreeConstraint}) require the vacuum value of the dilaton to lie at the $\bZ_3$ point, $\phi  = \exp( \pi i /3)$. 

The possibility of enlarging the R-symmetry for special values of the fluxes will in fact be realized in all of our explicit Fermat models, as we shall tabulate in section~\ref{ExampleSec}.

\subsection{Counting $W=0$ vacua}

As described in section 2, the set of all IIB flux vacua satisfying $DW=0$ with tadpole no greater than $L$ scales like $\sqrt{L}^{2 b_3}$, where $2 b_3$ is the number of fluxes.  One is interested in the counting of vacua obeying $W=0$ as well.

In \cite{DGKT}, an estimate was made for the number of $W=0$ vacua, \begin{eqnarray}
\label{OldCounting}
N_{\rm vacua} \sim \sum_{H=1}^{H_{\rm max}} H^{2 {\cal D} - \eta} L^{b_3 - (\eta + {\cal D})/2} \,,
\end{eqnarray}
with in our notation ${\cal D} \equiv \dim {\cal V}$, $\eta = b_3\, \dim {\cal V}/2$.  Here the {\it height} $H(\phi)$ takes into account the value of the dilaton appearing in the $D_{\varphi^A} W  = 0$ equations and its effect on the ability to solve the equations by varying the fluxes.

However, although the formula (\ref{OldCounting}) correctly accounts for the dilaton dropping out of the equations $D_\phi W = W = 0$, it does not take into account the resulting triviality and dilaton-independence of the $D_{\varphi^A} W = 0$ equations.  Hence for the class of vacua we study, (\ref{OldCounting}) will not give a proper counting of vacua; the dilaton-independence means the height $H(\phi)$ does not contribute, and the triviality of the $D_{\varphi^A} W = 0$ equations means fewer constraints are imposed.

Instead, the counting of vacua is much simpler.  The dimension of the space of solutions for each $f$ and $h$ is simply the number of eigenvalues of $A$ that unfaithfully represent $\bZ_d$, as well as being compatible with the periodicity $D$.  This is equivalent to imposing $\dim {\cal V}$ constraints on the $b_3$ available fluxes.  Taking into account both $f$ and $h$ flux, we find the counting
\begin{eqnarray}
\label{CountVacua}
N_{\rm vacua}(W=0) \sim L^{b_3 - \dim {\cal V}} \,,
\end{eqnarray}
with $b_3 - \dim {\cal V}$ the number of acceptable eigenvalues.
Recalling that all flux vacua scale as $L^{b_3}$ (\ref{AllVacScaling}), we find that the suppression is given precisely by the dimension of the vector space of periods,
\begin{eqnarray}
\label{SupFactor}
{N_{\rm vacua}(W=0) \over N_{\rm vacua} } \sim L^{- \dim {\cal V}} \,.
\end{eqnarray}
One would like to estimate (\ref{CountVacua}) as $b_3$ grows.  In particular, do $W=0$ vacua make up a sizable fraction of all vacua in this limit, or become a set of measure zero?

We recall from (\ref{Inequality}) that both $b_3$ and $\dim {\cal V}$ lie between $\phi(d)$ and $d$.  Hence it is useful to understand the behavior of the Euler totient function $\phi(d)$ as $d$ grows; the following results may be found, for example, in \cite{Totient}.  This function does not behave smoothly, as one has $\phi(d) = d-1$ for $d$ prime, while it may be considerably smaller han $d$ when the integer is highly composite.  One always has $\phi(d) \geq \sqrt{d}$ except for $d=2,6$, and for $d$ not prime there is an upper bound:
\begin{eqnarray}
\sqrt{d} \leq \phi(d) \leq d - \sqrt{d} \,, \quad \quad d \; {\rm composite} \,.
\end{eqnarray}
Models where the suppression factor (\ref{SupFactor}) is least for fixed $b_3$ are hence those where $b_3 \sim d$ and $\dim {\cal V} \sim \phi(d) \sim  \sqrt{d}$, giving
\begin{eqnarray}
{N_{\rm vacua}(W=0) \over N_{\rm vacua} } \sim L^{-\sqrt{b_3}} \,.
\end{eqnarray}
Hence, as $b_3$ grows large, $W=0$ vacua are invariably more and more suppressed; however, for these least suppressed cases the suppression is a strictly smaller power than the total number of vacua, indicating that the number of $W=0$ vacua to leading order in $b_3$ scales the same as the total number of vacua, going as $L^{b_3}$.

Furthermore, for $d \to \infty$ the totient function tends towards
\begin{eqnarray}
\phi(d \to \infty) \rightarrow e^{-\gamma} \, {d \over \log \log d} \,,
\end{eqnarray}
where $\gamma$ is the 
Euler-Mascheroni constant with $e^{-\gamma} \approx 0.561$.  Hence for all $b_d \sim d$, $\dim {\cal V} \sim \phi(d)$ models the suppression factor (\ref{SupFactor}) remains strictly smaller than the total number of vacua, and again to leading order in $b_3$ the number of $W=0$ vacua is $L^{b_3}$.

Making more precise statements requires being able to determine more precisely how the dimension of the vector space of periods $\dim {\cal V}$ differs from $\phi(d)$ in the limit of large $b_3$.  In particular, in making the above estimates we have assumed that taking $\dim {\cal V} \sim \phi(d)$ is valid for some models in this limit.
We do not analyze this further here, but leave it as an interesting open question.

\section{Examples}
\label{ExampleSec}

In this section we apply the analysis of the previous sections to some simple hypersurfaces in weighted projective space: the Fermat models with one, two and three parameters enumerated in \cite{SimpleModels}.  The one-parameter cases were already described in \cite{DGKT} and our results agree with that discussion.  We see that the case for the two- and three-parameter models are similar: not every space is compatible with $W=0$ vacua, but a significant fraction are. 

We also describe a non-Fermat model possessing $W=0$ vacua.  At the end of the section we discuss generalizations of these results, and present an example of a Calabi-Yau defined as the intersection of multiple polynomials.

\subsection{Fermat models}

\begin{table}
\begin{center}
\begin{tabular}{|c|c|c|c|c|} \hline 
$(k_1, k_2, k_3, k_4, k_5)$ & $d$ & $\{ n_J \}$ & $\phi(d)$  & R-sym \\ \hline \hline
$(1,1,1,1,1)$ &$5$ & $\{ 1,2,3,4 \}$ &$4$ & ---\\ \hline
$(1,1,1,1,2)$ &$6$ & $\{ 1,{\bf 2}, {\bf 4},5 \}$ &$2$ & $\bZ_2 (\bZ_6)$ \\ \hline
$(1,1,1,1,4)$ &$8$ & $\{ 1,3,5,7 \}$ &$4$ & ---\\ \hline
$(1,1,1,2,5)$ &$10$ & $\{ 1,3,7,9 \}$ &$4$ & ---\\ \hline
\end{tabular}
\end{center}
\caption{One-parameter models ($b_3 = 4$).
\label{TableOne}}
\end{table}

In one-parameter models the $U(\varphi^A)$ are absent and the periodicity condition is therefore absent as well.  Hence the only question is whether the R-symmetry constraint can be imposed, or equivalently, whether $b_3 - \phi(d) > 0$, which is the language in which the issue was studied in \cite{DGKT}.  Of the four models listed in table \ref{TableOne}, only the $d=6$ model $(1,1,1,1,2)$ satisfies this relation.  

Besides the values of $d$ and $\phi(d)$ for each model, we have listed the period indices $\{n_J\}$, which give the eigenvalues $\alpha^{n_J}$ of the monodromy matrix $A$, with $\alpha^d = 1$.  We have denoted in bold those $b_3 - \phi(d)$ eigenvalues that share a common factor with $d$ and hence do not faithfully represent $\bZ_d$.  The last column displays the R-symmetry for the model that does possess $W=0$ vacua, and in parentheses the enlarged symmetry present when combined with a transformation of the dilaton; those with no R-symmetry do not have $W=0$ vacua.
This analysis is completely in agreement with \cite{DGKT}.

We now turn to the two- and three-parameter hypersurfaces as listed in \cite{SimpleModels} (we leave out a three-parameter model that has a non-monomial deformation).  Tables \ref{TableTwo} and \ref{TableThree} contain additional columns listing the monomials corresponding to the $\varphi^A$ moduli, as well as the total periodicity $D$ of those monomials.  Again those eigenvalues of $A$ that are not faithful are bold, and when $W=0$ vacua are present, the R-symmetry is listed.
The $W=0$ vacua for the model $(1,1,1,2,6)$ were previously found in \cite{GKTT}.

\begin{table}
\begin{center}
\begin{tabular}{|c|c|c|c|c|c|c|} \hline 
$(k_1, k_2, k_3, k_4, k_5)$ & $d$ & $\{ n_J \}$ & $\phi(d)$ & monom & $D$ &R-sym \\ \hline \hline
$(1,1,2,2,2)$ &$8$ & $\{ 1,{\bf 2},3,5,{\bf 6},7 \}$ &$4$& $x_1^4 x_2^4$ &$2$ & $\bZ_4 (\bZ_8)$ \\ \hline
$(1,1,2,2,6)$ &$12$ & $\{ 1,{\bf 3},5,7,{\bf 9},11 \}$ &$4$ & $x_1^6 x_2^6$ & $2$ &$\bZ_6 (\bZ_{12})$ \\ \hline
$(1,3,2,2,4)$ &$12$ & $\{ 1,{\bf 2},5,7,{\bf 10},11 \}$ &$4$& $ x_1^6 x_2^2$ &$2$ &$\bZ_2 (\bZ_6)$ \\ \hline
$(1,7,2,2,2)$ &$14$ & $\{ 1,3,5,9,11,13 \}$ &$6$ & $x_1^7 x_2$ &$2$ &---\\ \hline
$(1,1,1,6,9)$ &$18$ & $\{ 1,5,7,11,13,17 \}$ &$6$ & $x_1^6 x_2^6 x_3^6$ &$3$& ---\\ \hline
\end{tabular}
\end{center}
\caption{Two-parameter models ($b_3 = 6$).
\label{TableTwo}}
\end{table}

\begin{table}
\begin{center}
\begin{tabular}{|c|c|c|c|c|c|c|} \hline 
$(k_1, k_2, k_3, k_4, k_5)$ & $d$ & $\{ n_J \}$ & $\phi(d)$ & monom & $D$ &R-sym \\ \hline \hline
$(1,2,3,3,3)$ &$12$ & $\{ 1,{\bf 2},{\bf 3},5,7,{\bf 9},{\bf 10},11 \}$ &$4$& $x_1^4 x_2^4, x_1^8 x_2^2$ &$3$ & $\bZ_4 (\bZ_{12})$ \\ \hline
$(1,3,3,3,5)$ &$15$ & $\{ 1,2,4,7,8,11,13,14 \}$ &$8$ & $x_1^5 x_5^2, x_1^{10} x_5$ & $3$ &--- \\ \hline
$(1,2,3,3,9)$ &$18$ & $\{ 1,{\bf 3},5,7,11,13,{\bf 15},17 \}$ &$6$& $ x_1^6 x_2^6, x_1^{12} x_2^3$ &$3$ &$\bZ_6 (\bZ_{18})$ \\ \hline
$(1,1,2,8,12)$ &$24$ & $\{ 1,5,7,11,13,17,19,23\}$ &$8$ & $x_1^6 x_2^6 x_3^6, x_1^{12} x_2^{12}$ &$4$ &---\\ \hline
\end{tabular}
\end{center}
\caption{Three-parameter models ($b_3 = 8$).
\label{TableThree}}
\end{table}

In most of these cases, all the eigenvalues of $A$ are compatible with the periodicity condition (\ref{PeriodCond}) set by $D$, namely that $N_J \equiv d/ \gcd(n_J, d) = \ell_J D$ for some integer $\ell_J$.   The exception is the $(1,2,3,3,3)$ model, which has four unfaithful eigenvalues, $\alpha^2, \alpha^3, \alpha^9, \alpha^{10}$ with $\alpha^{12} = 1$.  For $\alpha^2$ and $\alpha^{10}$, we can satisfy the condition $g = g \cdot A^6$, compatble with the periodicity $D=3$.  For the other two eigenvalues, however, the corresponding condition would be $g = g \cdot A^4$, which is incompatbible with $D=3$; hence only the first two eigenvalues generate acceptable flux vector solutions.  This corresponds precisely to the fact that in this example, $\dim {\cal V} > \phi(d)$ as discussed in section \ref{ConstructSec}; we have $\dim {\cal V} = 6$ while $\phi(d) = 4$.  The dimension of available fluxes is then $b_3 - \dim {\cal V} = 2$, rather than $b_3 - \phi(d) = 4$.

Hence we see for these simple examples that $W=0$ vacua, while not present in every model, are not uncommon and arise in an order one fraction of the cases considered.  

\subsection{Non-Fermat models}

Most hypersurfaces in weighted projective space are not Fermat, so it is useful to identify non-Fermat models for which this construction applies.  Of the sixteen non-Fermat models with two or three parameters listed in \cite{BKK, HLY} we find two cases with $W=0$ vacua.  The simpler is the two-parameter model with weights $(1,1,1,2,3)$ and $d=8$.  Here $b_3 = 6$, $\phi(8) = 2$ and the $b_3 - \phi(8) = 2$ unfaithful eigenvalues are $\alpha^2, \alpha^6$ with $\alpha^8=1$, allowing us to impose the R-symmetry constraint
\begin{eqnarray}
\label{NonFermatR}
g = g \cdot A^4\,.
\end{eqnarray}
To calculate the periodicity $D$, we proceed as follows.  A choice for the defining polynomial is
\begin{eqnarray}
P_0(x_i) = x_1^8 + x_2^8 + x_3^8 + x_4^4 + x_5^2 x_4 \,,
\end{eqnarray}
leading to the mirror polynomial
\begin{eqnarray}
\hat{P}_0(y_i) = y_1^8 + y_2^8 + y_3^8 + y_4^4 y_5 + y_5^2 \,,
\end{eqnarray}
defined in $\wcp^4_{1,1,1,1,4}$ with $\hat{d} = 8$.  The orbifold group defining the mirror is $H = \bZ_8 \times \bZ_8$, and the monomial other than $y_1 y_2 y_3 y_4 y_5$ invariant under $H$ is $(y_1 y_2 y_3 y_4)^2$.  Hence $Q =2$ for this monomial, and the periodicity is $D=d/Q = 4$.  Since $N=4=D$ in (\ref{NonFermatR}), the periodicity condition is satisfied and we 
indeed find $W=0$ vacua in this example; the R-symmetry is
\begin{eqnarray}
\bZ_4: \quad \psi \to \alpha^2 \psi, \quad \varphi \to - \varphi \,. 
\end{eqnarray}
The three-parameter model $(1,1,2,3,5)$ with $d=12$ has $W=0$ vacua as well, as one can show in an analogous fashion; there only two of the four unfaithful eigenvalues are compatible with the periodicity condition.

Thus we see that $W=0$ vacua are not restricted to the Fermat cases.  In this small sample the Fermat cases seem to have a higher incidence of $W=0$ vacua, but one should bear in mind 
that the non-Fermat class always includes those models with prime $d$ (except the quintic), which never support $W=0$ vacua by this construction.  It would be interesting to quantify more precisely the statement of the periodicity condition for non-Fermat examples, and thus understand whether $W=0$ vacua are of greater (or lesser) incidence in Fermat cases than in non-prime non-Fermat models.

\subsection{Complete intersection Calabi-Yau models}

Hypersurfaces in weighted projective space represent a subclass of the more general class of spaces defined as the intersections of multiple hypersurfaces in products of projective spaces.  We have not discussed this more general class at all, but one naturally wonders whether our methods may be applied in that more general context.  

In this section we give an example of an analogous construction of $W=0$ vacua in a simple space  \cite{Berglundetal} defined via the vanishing of two cubic hypersurfaces in $\cp^5$:
\begin{eqnarray}
P_1 = x_1 x_2 x_3 - 3 \psi x_4 x_5 x_6 \,, \quad P_2 = x_4 x_5 x_6 - 3 \psi x_1 x_2 x_3 \,,
\end{eqnarray}
where the single parameter $\psi$ is the only one surviving the $H$-projection to the mirror \cite{LT}.  There is a $\bZ_6$ monodromy around $\psi = 0$, represented on a Picard-Fuchs basis of periods as
\begin{eqnarray}
\label{AMatrixCICY}
A = \begin{pmatrix}
-4\; -3\; -2 \;\;\; 1 \; \cr 
\;\;1\;\;\;\;\; 0\;\;\;\;\; 0 \;\;\;\; 0 \; \cr 
\;\;0\;\;\;\;\; 1\;\;\;\;\; 0 \;\;\;\;0 \; \cr 
\;-8\; -6\; -5 \;\;\; 2 \;\;\; \cr 
\end{pmatrix}\,.
\end{eqnarray}
This transformation has the interesting property that $A^6 \neq 1$ (in fact there is no power of $A$ that is the identity); this is related to a logarithmic singularity in the periods at $\psi = 0$, absent for simple hypersurfaces.  Nonetheless, the eigenvalues of $A$ do represent $\bZ_6$; in fact they represent $\bZ_6$ unfaithfully, as they are all cube roots of unity.  Hence we can impose the R-symmetry constraint
\begin{eqnarray}
f = f\cdot A^3 \,, \quad h = h \cdot A^3 \,,
\end{eqnarray}
to obtain fluxes that lead to $W=0$ vacua.\footnote{$W=0$ vacua in this model were independently found by A.~Giryavets.}  The associated R-symmetry is
\begin{eqnarray}
\bZ_2: \quad \psi \to - \psi \,.
\end{eqnarray}
Since all four eigenvalues of $A^3$ are $+1$, naively one might think that {\it all} fluxes produce $W=0$ vacua; however, two of the associated eigenvectors of $A^3$ actually vanish, a novel occurance for the CICY models that we did not encounter previously.  Hence the $W=0$ vacua occur at codimension two in the space of fluxes.  As this is a one-parameter model, there is no periodicity condition; the logarithmic singularity of the periods at $\psi  = 0$ is removed by a K\"ahler transformation.

For the other one-parameter CICY models presented in \cite{Berglundetal}, there is always a $\bZ_d$ monodromy action around $\psi =0$, with $d$ the sum of the degrees of the polynomials.  A similar analysis then holds for them.  We have not considered more complicated CICY spaces in any further detail, but we see no reason in principle why this construction of $W=0$ vacua does not apply over the entire class.

\section{Conclusions}
\label{ConclusionSec}

We have demonstrated a general construction for obtaining type IIB flux vacua with vanishing superpotential and discrete R-symmetries.  The discrete R-symmetry appears at fixed loci of the corresponding modular transformation, in particular at the Landau-Ginzburg point on the $\psi$-plane, when the monodromy action on the periods possesses eigenvalues {\it unfaithfully} representing the monodromy group.  The remaining condition that these eigenvalues are compatible with the periodicity of the additional complex structure parameters can be thought of as a requirement that these moduli are at fixed points of the modular transformation as well.
We formulated our construction for hypersurfaces in weighted projective space, but we see no reason why it should not generalize to the broader class of complete intersections in products of projective spaces, and we gave a simple example in section~\ref{ExampleSec}.

We have also showed that the counting (as usual at large charge tadpole $L$) of the $W=0$ vacua in the most favorable models approaches the same power law as the set of all vacua as the number of moduli grows large; the ratio is subleading in $b_3$.  There are also many spaces with no $W=0$ vacua.  Understanding this distribution better involves an improved understanding of the asymptotic behavior of the vector space of periods.

The nature of the construction, that it is valid over a range of parameters, means that not all complex structure moduli are fixed by the fluxes; this is in addition to the K\"ahler moduli, which are never stabilized by fluxes in a IIB vacuum.  The nonperturbative effects \cite{KKLT} that can stabilize K\"ahler moduli in general also depend on the complex structure moduli, so it is conceivable that all moduli could still be stabilized once all corrections are taken into account.

We make no claim that our construction exhausts all $W=0$ vacua.  It is possible that vacua exist at isolated points, rather than only over a continuous range as we show here.  The rank of the vector space of periods ${\cal V}$ can become smaller at subloci on moduli space, as can the ranks of the vector spaces ${\cal V}^{(n)}$ associated with derivatives of the periods; if this happens all at the same point isolated vacua with vanishing superpotential could arise.

This study was motivated partially by the findings of \cite{DGKT} that $W=0$ vacua in one-parameter models were associated with particular arithmetic structure, with the periods living in field extensions of the rational numbers of small degree, in particular cyclotomic fields for the hypersurfaces in weighted projective space.  We find here, however, that in general $W=0$ vacua are {\it not} associated with an arithmetic structure.  The place of the field extension is taken by the ``vector space of periods", which as the name implies maintains the vector space properties of the extensions without preserving an obvious product.  Special points in moduli space exist at least in some cases where the vector space reduces in dimension, and may be promoted to a field, associated with attractor points and complex multiplication \cite{AA}.  It seems likely that further study of these connections could shed more light on the mathematical structure of the periods and the associated vacua.

\section*{Acknowledgments}

I benefited from discussions and correspondence with M.~Dine, A.~Giryavets, S.~Gukov, S.~Kachru, S.~Katz, D.~Morrison, Z.~Sun and W.~Taylor.  This work was supported by NSF grant PHY-0243680. Any opinions, findings, and conclusions or recommendations expressed in this material are those of the author and do not necessarily reflect the views of the National Science Foundation.


\begin{thebibliography}{10}
\baselineskip=15pt


\bibitem{Heterotic}
 E.~I.~Buchbinder and B.~A.~Ovrut,
  ``Vacuum stability in heterotic M-theory,''
  Phys.\ Rev.\ D {\bf 69}, 086010 (2004)
  [arXiv:hep-th/0310112];
S.~Gukov, S.~Kachru, X.~Liu and L.~McAllister,
  ``Heterotic moduli stabilization with fractional Chern-Simons invariants,''
  Phys.\ Rev.\ D {\bf 69}, 086008 (2004)
  [arXiv:hep-th/0310159];
R.~Brustein and S.~P.~de Alwis,
  ``Moduli potentials in string compactifications with fluxes: Mapping the
  discretuum,''
  Phys.\ Rev.\ D {\bf 69}, 126006 (2004)
  [arXiv:hep-th/0402088];
  K.~Becker, M.~Becker, P.~S.~Green, K.~Dasgupta and E.~Sharpe,
  ``Compactifications of heterotic strings on non-Kaehler complex  manifolds.
  II,''
  Nucl.\ Phys.\ B {\bf 678}, 19 (2004)
  [arXiv:hep-th/0310058];
  M.~Becker, G.~Curio and A.~Krause,
  ``De Sitter vacua from heterotic M-theory,''
  Nucl.\ Phys.\ B {\bf 693}, 223 (2004)
  [arXiv:hep-th/0403027];
S.~Gurrieri, A.~Lukas and A.~Micu,
  ``Heterotic on half-flat,''
  Phys.\ Rev.\ D {\bf 70}, 126009 (2004)
  [arXiv:hep-th/0408121];

\bibitem{MTheory}

K.~Dasgupta, G.~Rajesh and S.~Sethi,
  ``M theory, orientifolds and G-flux,''
  JHEP {\bf 9908}, 023 (1999)
  [arXiv:hep-th/9908088];
 B.~S.~Acharya,
  ``A moduli fixing mechanism in M theory,''
  arXiv:hep-th/0212294;
B.~de Carlos, A.~Lukas and S.~Morris,
  ``Non-perturbative vacua for M-theory on G(2) manifolds,''
  JHEP {\bf 0412}, 018 (2004)
  [arXiv:hep-th/0409255];
B.~S.~Acharya, F.~Denef and R.~Valandro,
  ``Statistics of M theory vacua,''
  arXiv:hep-th/0502060.
P.~S.~Aspinwall,
  ``An analysis of fluxes by duality,''
  arXiv:hep-th/0504036;
P.~S.~Aspinwall and R.~Kallosh,
  ``Fixing all moduli for M-theory on K3 x K3,''
  arXiv:hep-th/0506014.


\bibitem{IIA}
 K.~Behrndt and M.~Cvetic,
  ``General N = 1 supersymmetric flux vacua of (massive) type IIA string
  theory,''
  arXiv:hep-th/0403049;
K.~Behrndt and M.~Cvetic,
  ``General N = 1 supersymmetric fluxes in massive type IIA string theory,''
  Nucl.\ Phys.\ B {\bf 708}, 45 (2005)
  [arXiv:hep-th/0407263];
J.~P.~Derendinger, C.~Kounnas, P.~M.~Petropoulos and F.~Zwirner,
  ``Superpotentials in IIA compactifications with general fluxes,''
  Nucl.\ Phys.\ B {\bf 715}, 211 (2005)
  [arXiv:hep-th/0411276];
S.~Kachru and A.~K.~Kashani-Poor,
  ``Moduli potentials in type IIA compactifications with RR and NS flux,''
  JHEP {\bf 0503}, 066 (2005)
  [arXiv:hep-th/0411279];
  D.~Lust and D.~Tsimpis,
  ``Supersymmetric AdS(4) compactifications of IIA supergravity,''
  JHEP {\bf 0502}, 027 (2005)
  [arXiv:hep-th/0412250];
T.~W.~Grimm and J.~Louis,
  ``The effective action of type IIA Calabi-Yau orientifolds,''
  Nucl.\ Phys.\ B {\bf 718}, 153 (2005)
  [arXiv:hep-th/0412277];
   G.~Villadoro and F.~Zwirner,
  ``N = 1 effective potential from dual type-IIA D6/O6 orientifolds with
  general fluxes,''
  arXiv:hep-th/0503169;
J.~P.~Derendinger, C.~Kounnas, P.~M.~Petropoulos and F.~Zwirner,
  ``Fluxes and gaugings: N = 1 effective superpotentials,''
  arXiv:hep-th/0503229;
O.~DeWolfe, A.~Giryavets, S.~Kachru and W.~Taylor,
  ``Type IIA moduli stabilization,''
  arXiv:hep-th/0505160;
  T.~House and E.~Palti,
  ``Effective action of (massive) IIA on manifolds with SU(3) structure,''
  arXiv:hep-th/0505177;
P.~G.~Camara, A.~Font and L.~E.~Ibanez,
  ``Fluxes, moduli fixing and MSSM-like vacua in a simple IIA orientifold,''
  arXiv:hep-th/0506066;
  J.~Bovy, D.~Lust and D.~Tsimpis,
  ``N = 1,2 supersymmetric vacua of IIA supergravity and SU(2) structures,''
  arXiv:hep-th/0506160.



\bibitem{IIB}

S.~B.~Giddings, S.~Kachru and J.~Polchinski,
  ``Hierarchies from fluxes in string compactifications,''
  Phys.\ Rev.\ D {\bf 66}, 106006 (2002)
  [arXiv:hep-th/0105097];
S.~Kachru, M.~B.~Schulz and S.~Trivedi,
  ``Moduli stabilization from fluxes in a simple IIB orientifold,''
  JHEP {\bf 0310}, 007 (2003)
  [arXiv:hep-th/0201028];
A.~R.~Frey and J.~Polchinski,
  ``N = 3 warped compactifications,''
  Phys.\ Rev.\ D {\bf 65}, 126009 (2002)
  [arXiv:hep-th/0201029];
     O.~DeWolfe and S.~B.~Giddings,
  ``Scales and hierarchies in warped compactifications and brane worlds,''
  Phys.\ Rev.\ D {\bf 67}, 066008 (2003)
  [arXiv:hep-th/0208123];
     P.~K.~Tripathy and S.~P.~Trivedi,
  ``Compactification with flux on K3 and tori,''
  JHEP {\bf 0303}, 028 (2003)
  [arXiv:hep-th/0301139];
  M.~Berg, M.~Haack and B.~Kors,
  ``An orientifold with fluxes and branes via T-duality,''
  Nucl.\ Phys.\ B {\bf 669} (2003) 3
  [arXiv:hep-th/0305183];
  M.~Berg, M.~Haack and B.~Kors,
  ``Brane / flux interactions in orientifolds,''
  Fortsch.\ Phys.\  {\bf 52} (2004) 583
  [arXiv:hep-th/0312172];
V.~Balasubramanian and P.~Berglund,
  ``Stringy corrections to Kaehler potentials, SUSY breaking, and the
  cosmological constant problem,''
  JHEP {\bf 0411}, 085 (2004)
  [arXiv:hep-th/0408054];
V.~Balasubramanian, P.~Berglund, J.~P.~Conlon and F.~Quevedo,
  ``Systematics of moduli stabilisation in Calabi-Yau flux compactifications,''
  JHEP {\bf 0503}, 007 (2005)
  [arXiv:hep-th/0502058].

\bibitem{KKLT}
S.~Kachru, R.~Kallosh, A.~Linde and S.~P.~Trivedi,
  ``De Sitter vacua in string theory,''
  Phys.\ Rev.\ D {\bf 68}, 046005 (2003)
  [arXiv:hep-th/0301240].
  
\bibitem{Nonpert}
F.~Denef, M.~R.~Douglas and B.~Florea,
  JHEP {\bf 0406}, 034 (2004)
  [arXiv:hep-th/0404257];
F.~Denef, M.~R.~Douglas, B.~Florea, A.~Grassi and S.~Kachru,
  ``Fixing all moduli in a simple F-theory compactification,''
  arXiv:hep-th/0503124.

\bibitem{CosConst}
J.~D.~Brown and C.~Teitelboim,
  ``Dynamical Neutralization Of The Cosmological Constant,''
  Phys.\ Lett.\ B {\bf 195}, 177 (1987);
J.~D.~Brown and C.~Teitelboim,
  ``Neutralization Of The Cosmological Constant By Membrane Creation,''
  Nucl.\ Phys.\ B {\bf 297}, 787 (1988);
R.~Bousso and J.~Polchinski,
  ``Quantization of four-form fluxes and dynamical neutralization of the
  cosmological constant,''
  JHEP {\bf 0006}, 006 (2000)
  [arXiv:hep-th/0004134];
J.~L.~Feng, J.~March-Russell, S.~Sethi and F.~Wilczek,
  ``Saltatory relaxation of the cosmological constant,''
  Nucl.\ Phys.\ B {\bf 602}, 307 (2001)
  [arXiv:hep-th/0005276].


\bibitem{Statistics}
M.~R.~Douglas,
  ``The statistics of string / M theory vacua,''
  JHEP {\bf 0305}, 046 (2003)
  [arXiv:hep-th/0303194].
  S.~Ashok and M.~R.~Douglas,
  ``Counting flux vacua,''
  JHEP {\bf 0401}, 060 (2004)
  [arXiv:hep-th/0307049];
F.~Denef and M.~R.~Douglas,
  ``Distributions of flux vacua,''
  JHEP {\bf 0405}, 072 (2004)
  [arXiv:hep-th/0404116];
  A.~Giryavets, S.~Kachru and P.~K.~Tripathy,
  ``On the taxonomy of flux vacua,''
  JHEP {\bf 0408}, 002 (2004)
  [arXiv:hep-th/0404243];
  J.~P.~Conlon and F.~Quevedo,
  ``On the explicit construction and statistics of Calabi-Yau flux vacua,''
  JHEP {\bf 0410}, 039 (2004)
  [arXiv:hep-th/0409215];
  J.~Kumar and J.~D.~Wells,
``Landscape cartography: A coarse survey of gauge group rank and
  stabilization of the proton,''
  Phys.\ Rev.\ D {\bf 71}, 026009 (2005)
  [arXiv:hep-th/0409218];
R.~Blumenhagen, F.~Gmeiner, G.~Honecker, D.~Lust and T.~Weigand,
  ``The statistics of supersymmetric D-brane models,''
  Nucl.\ Phys.\ B {\bf 713}, 83 (2005)
  [arXiv:hep-th/0411173];
    F.~Denef and M.~R.~Douglas,
  ``Distributions of nonsupersymmetric flux vacua,''
  JHEP {\bf 0503}, 061 (2005)
  [arXiv:hep-th/0411183];
J.~Gomis, F.~Marchesano and D.~Mateos,
  ``An open string landscape,''
  arXiv:hep-th/0506179.


\bibitem{DGT}
M.~Dine, E.~Gorbatov and S.~Thomas,
  ``Low energy supersymmetry from the landscape,''
  arXiv:hep-th/0407043.




\bibitem{Scales}
L.~Susskind,
  ``Supersymmetry breaking in the anthropic landscape,''
  arXiv:hep-th/0405189.
  M.~R.~Douglas,
  ``Statistical analysis of the supersymmetry breaking scale,''
  arXiv:hep-th/0405279.


\bibitem{DOS}
 M.~Dine, D.~O'Neil and Z.~Sun,
  ``Branches of the landscape,''
  arXiv:hep-th/0501214;
  M.~Dine,
  ``The intermediate scale branch of the landscape,''
  arXiv:hep-th/0505202.


\bibitem{DGKT}
O. DeWolfe, A. Giryavets, S. Kachru and W. Taylor, ``Enumerating flux
vacua with enhanced symmetries,'' JHEP {\bf 0502}
(2005) 037 [arXiv:hep-th/0411061].



\bibitem{AA}
G.~W.~Moore,
  ``Arithmetic and attractors,''
  arXiv:hep-th/9807087.

\bibitem{OtherGeom}
C.~Borcea, ``Calabi-Yau Threefolds and Complex Multiplication,''
in {\it Essays on Mirror Manifolds}, S.T. Yau ed., International Press, 1992;
S.~Gukov and C.~Vafa,
  ``Rational conformal field theories and complex multiplication,''
  Commun.\ Math.\ Phys.\  {\bf 246}, 181 (2004)
  [arXiv:hep-th/0203213];
G.~W.~Moore,
  ``Les Houches lectures on strings and arithmetic,''
  arXiv:hep-th/0401049;
P.~Candelas, X.~de la Ossa and F.~Rodriguez-Villegas,
  ``Calabi-Yau manifolds over finite fields. I,''
  arXiv:hep-th/0012233;
P.~Candelas, X.~de la Ossa and F.~Rodriguez Villegas,
  ``Calabi-Yau manifolds over finite fields. II,''
  arXiv:hep-th/0402133.
  
\bibitem{DS}
M.~Dine and Z.~Sun, ``R symmetries in the landscape,"
arXiv:hep-th/0506246.


\bibitem{GVW}
S.~Gukov, C.~Vafa and E.~Witten,
  ``CFT's from Calabi-Yau four-folds,''
  Nucl.\ Phys.\ B {\bf 584}, 69 (2000)
  [Erratum-ibid.\ B {\bf 608}, 477 (2001)]
  [arXiv:hep-th/9906070];
T.~R.~Taylor and C.~Vafa,
  ``RR flux on Calabi-Yau and partial supersymmetry breaking,''
  Phys.\ Lett.\ B {\bf 474}, 130 (2000)
  [arXiv:hep-th/9912152].

\bibitem{Berglundetal}
P.~Berglund, P.~Candelas, X.~De La Ossa, A.~Font, T.~H\"ubsch, D.~Jancic and F.~Quevedo,
  ``Periods for Calabi-Yau and Landau-Ginzburg vacua,''
  Nucl.\ Phys.\ B {\bf 419}, 352 (1994)
  [arXiv:hep-th/9308005].

  \bibitem{BKK}
P.~Berglund, S.~Katz and A.~Klemm,
  ``Mirror symmetry and the moduli space for generic hypersurfaces in toric
  varieties,''
  Nucl.\ Phys.\ B {\bf 456}, 153 (1995)
  [arXiv:hep-th/9506091].

  
\bibitem{GreenePlesser}
  B.~R.~Greene and M.~R.~Plesser,
  ``Duality In Calabi-Yau Moduli Space,''
  Nucl.\ Phys.\ B {\bf 338}, 15 (1990).

\bibitem{BerglundHubsch}
P.~Berglund and T.~H\"ubsch,
  ``A generalized construction of mirror manifolds,''
  Nucl.\ Phys.\ B {\bf 393}, 377 (1993)
  [arXiv:hep-th/9201014].

\bibitem{GKTT}
A.~Giryavets, S.~Kachru, P.~K.~Tripathy and S.~P.~Trivedi,
  ``Flux compactifications on Calabi-Yau threefolds,''
  JHEP {\bf 0404}, 003 (2004)
  [arXiv:hep-th/0312104].
  
  \bibitem{Candelasetal}
P.~Candelas, X.~C.~De La Ossa, P.~S.~Green and L.~Parkes,
  ``A Pair Of Calabi-Yau Manifolds As An Exactly Soluble Superconformal
  Theory,''
  Nucl.\ Phys.\ B {\bf 359}, 21 (1991).

  
\bibitem{BK}
P.~Berglund and S.~Katz,
  ``Mirror symmetry for hypersurfaces in weighted projective space and
  topological couplings,''
  Nucl.\ Phys.\ B {\bf 420}, 289 (1994)
  [arXiv:hep-th/9311014].



\bibitem{Morrison}
D.~R.~Morrison, ``Making enumerative predictions," Mirror Symmetry II (B. Greene and S.-T. Yau, eds.), International Press, Cambridge, 1997, pp. 457-482.


\bibitem{Totient}
G.~H.~Hardy and E.~M.~Wright, {\it An Introduction to the Theory of Numbers}, Oxford University Press, 1938.


\bibitem{SimpleModels}
P.~Candelas, X.~De La Ossa, A.~Font, S.~Katz and D.~R.~Morrison,
  ``Mirror symmetry for two parameter models. I,''
  Nucl.\ Phys.\ B {\bf 416}, 481 (1994)
  [arXiv:hep-th/9308083];
  S.~Hosono, A.~Klemm, S.~Theisen and S.~T.~Yau,
  ``Mirror symmetry, mirror map and applications to Calabi-Yau hypersurfaces,''
  Commun.\ Math.\ Phys.\  {\bf 167}, 301 (1995)
  [arXiv:hep-th/9308122];
P.~Candelas, A.~Font, S.~Katz and D.~R.~Morrison,
  ``Mirror symmetry for two parameter models. II,''
  Nucl.\ Phys.\ B {\bf 429}, 626 (1994)
  [arXiv:hep-th/9403187].


\bibitem{HLY}
S.~Hosono, B.~H.~Lian and S.~T.~Yau,
  ``GKZ generalized hypergeometric systems in mirror symmetry of Calabi-Yau
  hypersurfaces,''
  Commun.\ Math.\ Phys.\  {\bf 182}, 535 (1996)
  [arXiv:alg-geom/9511001].
  
\bibitem{LT}
A.~Libgober and J.~Teitelbaum,
  ``Lines on Calabi-Yau complete intersections, mirror symmetry, and
  Picard-Fuchs equations,''
  arXiv:alg-geom/9301001.



\end{thebibliography}
\end{document}